\documentstyle[twocolumn,pre,aps,epsf]{revtex}
\begin{document}

\title{
Role of entropy barriers for diffusion in the periodic potential}
\author{O.\ M.\ Braun\cite{Braun}}
\address{
Institute of Physics,
National Ukrainian Academy of Sciences,
03650 Kiev, Ukraine}
\maketitle

\begin{abstract}

Diffusion of a particle in the $N$-dimensional
external potential which is periodic 
in one dimension and unbounded 
in the other $N-1$ dimensions is investigated.
We find an analytical expression
for the overdamped diffusion
and study numerically
the cases of moderate and low damping.
We show that in the underdamped limit,
the multi-dimensional effects
lead to reduction (comparing with the one-dimensional motion)
of jump lengths between subsequent trapping
of the atom in bottoms
of the external periodic potential.
As application we consider the diffusion of a dimer
adsorbed on the crystal surface.

\end{abstract}

\section{Introduction}
\label{s1}

A variety phenomena in physics and other fields
can be modeled as Brownian motion
in an external periodic potential
\cite{Risken,Hanggi,Melnikov}.
One of particular example, the surface diffusion
of atoms or small clusters,
is of great fundamental and technological interest
\cite{DiffSurveys}.
During crystal growth the deposited atoms
diffuse over the surface until they become
incorporated in the lattice.
On the semiconductor Si(100) or Ge(100) surface,
most of the deposited Si or Ge atoms combine to form dimers,
and the diffusion of such dimers has recently been studied
experimentally by a scanning tunneling microscope
\cite{Harold1}.
Moreover, atoms adsorbed on metal surfaces
in some cases form closely packed islands
which diffuse as a whole
\cite{IslandsE,IslandsT}.

Theoretically the problem of diffusion can be described by
a Langevin equation for the atom(s) or, equivalently,
a Fokker-Planck-Kramers equation for
the distribution function in the phase space
\cite{Risken,Hanggi}.
In the trivial case of Brownian motion
without the external potential,
the diffusion coefficient is equal to
$D=D_f \equiv k_B T/m \eta$,
where
$k_B$ is the Boltzmann constant,
$T$ is the temperature,
$m$ is the particle mass, and
$\eta$ is the viscous damping coefficient
which models the energy exchange with 
the substrate (thermostat).
For a single atom in the one-dimensional (1D)
sinusoidal external (substrate) potential
the value of diffusion coefficient is well known
as summarized in the Risken monograph
\cite{Risken}.
Exact results exist
for the overdamped (Smoluchowski) case, $\eta \rightarrow \infty$,
when \cite{over}
\begin{equation}
D=D_f {\rm I}_0^{-2} \left( h \right),
\label{DHF}
\end{equation}
where
$h \equiv \varepsilon / 2 k_B T$,
$\varepsilon$ is the (total) height of the substrate potential and
${\rm I}_0 (h)$ is the modified Bessel function,
and
in the underdamped limit, $\eta \rightarrow 0$,
when \cite{Risken}
\begin{equation}
D=D_f G(h),
\label{DLF}
\end{equation}
where
$
G(h)=
\left( h / 2\pi \right)^{{1}/{2}}
e^h \, 
{\rm I}_0^{-1} \left( h \right) \,
J(h)
$, 
$
J(h) =
\int_0^1 du \;
u^{-{3}/{2}} \,
e^{-2h/u} \,
{\mathbf E}^{-1}(u)
$,
and ${\mathbf E}(u)$
is the complete elliptic integral of second kind.
At low temperatures,
$k_B T \ll \varepsilon$,
both expressions (\ref{DHF}) and (\ref{DLF})
take the Arrhenius form,
$D=\widetilde{D} A$ with 
$A=\exp \left( -\varepsilon / k_B T \right)$
and
$\widetilde{D} \approx 
\omega_0^2 a^2 / 2\pi \eta$
in the high-friction case
[here $a$ is the period of the substrate potential and
$\omega_0=(2\pi/a)(\varepsilon/2m)^{1/2}$ 
is the frequency of oscillation
at its minimum], and
$\widetilde{D} \approx \pi D_f /2$
in the low-friction limit.
In a general case the diffusion coefficient
can be found numerically
with practically any desired accuracy
by the matrix continued-fraction-expansion method
\cite{Risken}.

At low temperatures, $k_B T \ll \varepsilon$,
when the diffusion proceeds by uncorrelated
thermally activated jumps over the barrier
from one minimum of the external potential to another,
the diffusion coefficient may be presented as
$D = R A \langle \lambda^2 \rangle$,
where
$R A$ is the rate of escape from a minimum
of the external potential
and
$\langle \lambda^2 \rangle$ is the mean-square jump length.
For a moderate or large damping,
$\eta \agt \omega_0$,
when only jumps for one period $a$
of the external potential are possible,
one should take $\lambda=a$ and
$R=R_{\rm TST} B(\eta)$,
where
$R_{\rm TST} = \omega_0 /2\pi$
is the escape rate given by the transition state theory (TST)
\cite{Hanggi,TST},
and the factor
$B(\eta) = (z^2 +1)^{1/2} -z$
with
$z=\eta /2\omega_s$
provides an interpolation
between the TST and overdamped limits
as was found by Kramers \cite{Kramers}
[here $\omega_s$ is the ``saddle'' frequency
at the saddle point $x=x_s$,
near which the external potential has the form
$V(x) \approx \varepsilon - 
\frac{1}{2} m \omega_s^2 (x-x_s)^2$;
for the sinusoidal potential
$\omega_s = \omega_0$].
The underdamped limit,
$\eta \ll \omega_0$,
is qualitatively different:
in this case
$R \approx 2 \eta \varepsilon / \pi k_B T 
\propto \eta$
as was found firstly by Kramers \cite{Kramers},
but the average jump length diverges as
$\lambda \propto \eta^{-1}$,
thus this again leads to the dependence 
$R \lambda^2 \propto \eta^{-1}$
similarly to the overdamped case.
The occurrence of long jumps, $\lambda >a$,
has been observed in a number of experiments
on surface diffusion
\cite{DiffSurveys,LongJumps}.
The interval from low to moderate friction is covered
by the Mel'nikov-Meshkov formula
\cite{Melnikov2}
\begin{equation}
R_{\rm MM} \approx
R_{\rm TST}
\exp \left( \frac{1}{\pi} \int_0^{\infty} du \;
\frac{ \ln \left[ 1- e^{-\Delta \left(
u^2 +\frac{1}{4} \right)} \right] }
{u^2 +\frac{1}{4}} \right),
\label{}
\end{equation}
where
$\Delta = 8 h \eta / \omega_0$.
Thus, the whole interval of frictions may be
described by the interpolation formula
$R \approx R_{\rm MM} B(\eta)$,
which was checked numerically in \cite{Italy1}.
Combining this expression for $R$
with the numerically calculated values of $D$,
one can find the distribution of jump lengths
\cite{Italy2}.
Note that the widely used TST expression
$D \approx R_{\rm TST} A a^2$, where the diffusion coefficient
does not depend on the damping coefficient,
operates in fact for a narrow interval of frictions
close to the point $\eta \sim \omega_0$ only
(which, fortunately, often corresponds to 
experimental situations).

Although the described above results
for one-dimensional diffusion
are very important and often lead to reasonable estimations
for experimentally measured diffusion coefficients,
in real systems the motion always takes place
in a $N>1$ configurational space.
Indeed, even for diffusion of a single 
atom adsorbed on a crystal surface
$N=2$ at least.
Besides, the diffusing object may have
internal degrees of freedom.
Multi-dimensional effects modify both
the escape rate $R$ and the jump length $\lambda$.
The escape rate can be presented as
$R=R_{\rm 1D} F$,
where the coefficient $F$ is
known as the ``entropy factor''
\cite{Vineyard}.
The value of $F$ can be found with the help of
transition state theory
\cite{TST}
which yields
$F \approx 
\left( \Pi_i \omega_{0,i} \right)/
\left( \Pi_i \omega_{s,i} \right)$,
where
$\omega_{0,i}$ are the frequencies at the minimum and
$\omega_{s,i}$ are the ``saddle'' frequencies
for all degrees of freedom $i$
except the given one along the diffusion path.
In this approach $F$ can be interpreted as
$F=\exp (\Delta S /k_B)$, where
$\Delta S$ is the difference in entropy of
the saddle and minimum-energy configurations.
The entropy factor is often used to explain
the ``compensation effect''
\cite{DiffSurveys},
when at experiment one observes that
a decrease of the activation energy
(calculated as a slope of the Arrhenius plot of
$\ln D$ versus $T^{-1}$)
is compensated by decreasing of the prefactor.
As for the jump length,
while for $\eta \agt \omega_0$
it still is given by $\lambda =a$,
in the underdamped limit it is modified qualitatively
comparing with the one-dimensional case.
In the multi-dimensional space, the path connecting adjoining minima
of the external potential
may not coincide with the direction of
easy crossing at the saddle point.
Therefore, the probability of deactivation
during long jumps is enhanced,
leading to the reduction of jump length,
$\lambda < \lambda_{\rm 1D}$
\cite{Zhdanov,Haug,Chen}.
In particular, for the 2D-periodic substrate potential
with the square symmetry it was found numerically
\cite{Chen} that $D \propto \eta^{-0.5}$
which gives $\lambda \propto \eta^{-0.75}$.

The multi-dimensional effects are also important
in diffusion of molecules or small clusters:
even for diffusion in the 1D periodic potential
(e.g., along ``channels'' on furrowed or stepped surfaces)
one has for the dimer diffusion $N=2$ at least.
Diffusion of the dimer was studied numerically
by Vollmer \cite{Vollmer}
with the help of matrix continued-fraction-expansion technique.
The adiabatically slow motion of a linear molecule
in the 1D sinusoidal potential was analyzed in
\cite{SSmine}, where
the adiabatic trajectory was found for a general case.
This allowed to find the activation barriers
and the minimum-energy and saddle-state frequencies
and then to estimate the diffusion coefficient.

The aim of the present paper is to study the multi-dimensional effects
in diffusion processes.
We consider two typical examples:
motion of a single atom in a ``channel'' which is
periodic in one dimension and parabolic in others,
and diffusion of a dimer (two-atomic molecule)
in the 1D sinusoidal potential.
We find an analytical solution for the overdamped case
and analyze numerically the dependence of diffusion coefficient
on the damping constant $\eta$.
The numerical results were obtained
with the Verlet algorithm
by calculating the trajectory $x(t)$
and then splitting it into $N_{\rm tr}$ pieces,
each of the time duration $\tau$.
The diffusion coefficient was then calculated as
$D={\langle \Delta x^2 \rangle}/{2 \tau}$.

The paper is organized as follows.
In Sec.\ \ref{s2} we obtain the analytical expression
for the diffusion coefficient in the overdamped limit.
In Sec.\ \ref{s3} we analyze the case of pure entropic barriers.
In Sec.\ \ref{s4} the activated diffusion of a single atom is studied.
In Sec.\ \ref{s5} the diffusion of a dimer is described.
Finally, Sec.\ \ref{s6} concludes the paper.

\section{Overdamped limit}
\label{s2}

Consider a particle moving
in the $N$-dimensional external potential
$V_N(x,y_1, \ldots ,y_{N-1})$ which
is periodic in the $x$ direction,
\begin{equation}
V_N(x+a, \ldots ) = V_N(x, \ldots ),
\label{V2}
\end{equation}
and grows unboundently in the other $N-1$ dimensions,
\begin{equation}
V_N(x,y_1, \ldots ,y_{N-1}) \rightarrow 
\frac{1}{2} \, \omega_i^2(x) \, y_i^2 
\;\;\;\mbox{if}\;\;\;
y_i \rightarrow \pm \infty,
\label{V3}
\end{equation}
where $\omega_i(x) > 0$ for all $x$.

With presence of a viscous friction,
the particle motion should be diffusive at long-time scale.
The diffusion coefficient $D$ can be found with
the Einstein relation $D = T \mu$, 
where 
the mobility $\mu$ describes the proportionality
between the linear current $j$ and 
the infinitesimal external dc force $f$
which causes this current, $j = \mu f$.
Therefore, we have to consider the particle motion
in the external potential
\begin{equation}
V_f (x,y_1, \ldots ,y_{N-1}) = V_N - f x,
\label{V1}
\end{equation}
and then take the limit $f \rightarrow 0$.

In the overdamped case, when the friction coefficient $\eta$ 
is much larger than
the characteristic system frequencies,
the motion of the particle is described by the Smoluchowski equation
\begin{equation}
\frac{\partial W}{\partial t} +
\vec{\nabla} \cdot \vec{J} = 0, \;\;\;
\vec{J} = - \eta^{-1} (W \, \vec{\nabla} V_f + T \, \vec{\nabla} W),
\label{Smo}
\end{equation}
where $W(x,y_1, \ldots ,y_{N-1};t)$ is the distribution function,
$\vec{J}(x,y_1, \ldots ,y_{N-1};t)$ is the density of particle's current,
and  the particle mass 
and Boltzmann constant are put to unity,
$m=1$ and $k_B=1$.

For a steady state, Eq.\ (\ref{Smo}) takes the form
\begin{equation}
T \, \frac{\partial W}{\partial x} +
  W \, \frac{\partial V_f}{\partial x} =
  - \eta J_x
\label{Smo1}
\end{equation}
for the $x$ component,
and a similar form for other degrees of freedom.
The density $\vec{J}$ of the current
should satisfy the equation
\begin{equation}
\frac{\partial J_x}{\partial x} +
\sum_{i=1}^{N}
\frac{\partial J_{y_i}}{\partial y_i} = 0.
\label{JJ}
\end{equation}

To reduce notations, below we consider the case of
$N=2$ only; generalization to the $N>2$ case is trivial.
Let us introduce the one-dimensional density and current as
\begin{equation}
\rho(x) = 
  \int_{-\infty}^{+\infty} dy \; W(x,y),
\label{rho}
\end{equation}
\begin{equation}
j(x) = 
  \int_{-\infty}^{+\infty} dy \; J_x(x,y).
\label{jx}
\end{equation}
Owing to the condition (\ref{V3}), the current $j(x)$
does not depend on $x$,
\begin{equation}
\frac{d j(x)}{d x} =
  - J_{y}(x,+\infty) +
    J_{y}(x,-\infty)
  = 0,
\label{jx1}
\end{equation}
where we have used Eq.\ (\ref{JJ}).
Thus, integrating both parts of Eq.\ (\ref{Smo1}) over $y$,
we obtain the one-dimensional equation
\begin{equation}
T \, \frac{d \rho(x)}{d x} +
  \rho(x) \, \frac{d V_F(x)}{d x}
  = - \eta j,
\label{Smo2}
\end{equation}
where we introduced the potential
$V_F(x)$ defined by the equation
\begin{equation}
\frac{d V_F(x)}{d x} = \left[ \rho(x) \right]^{-1}
  \int_{-\infty}^{+\infty} dy \;
  W(x,y) \, \frac{\partial V_f(x,y)}{\partial x}.
\label{WF}
\end{equation}
Now, if $V_F(x)$ may be presented in the form
\begin{equation}
V_F(x) = V_N(x;f) - f x,
\label{WN}
\end{equation}
where $V_N(x;f)$ is a periodic function on $x$,
Eq.\ (\ref{Smo2}) takes the form studied in \cite{over},
and the diffusion coefficient can be calculated as
\begin{equation}
D = D_f (I_{+} I_{-})^{-1},
\;\;\;
I_{\pm}(T) = (2 \pi)^{-1} 
\int_0^{2 \pi} dx \;
e^{\pm V_{\rm eff}(x)/T},
\label{D0}
\end{equation}
where $D_f = T / \eta$ and
$V_{{\rm eff}}(x) = \lim_{f \rightarrow 0} V_N(x;f)$.
Thus, the diffusion coefficient $D$
is determined by the one-dimensional function $V_N(x;0)$.
In the limit $f \rightarrow 0$
we may substitute
the equilibrium distribution function
$W = W_{\rm eq} \propto \exp (-V_N /T)$
into Eq.\ (\ref{WF}), thus obtaining
\begin{equation}
\frac{d V_{{\rm eff}}(x)}{d x} =
  \frac{
  \int_{-\infty}^{+\infty} dy \;
   e^{-V_N(x,y)/T} \;
   \partial V_N(x,y) / \partial x
}{
  \int_{-\infty}^{+\infty} dy \;
  e^{-V_N(x,y)/T}
}.
\label{Veff1}
\end{equation}
Emphasize that this is the key approximation
which is rigorous in the overdamped limit only.
For the underdamped case, $\eta \rightarrow 0$,
a similar multiplicative separation
in the Fokker-Planck-Kramers equation,
$W(x,y,v_x,v_y,f) \propto
W(x,v_x,f) \, W_{\rm eq}(y,v_y)$,
does not work even in the $f \rightarrow 0$ limit.

Let $V_N (x,y)$ has the form
\begin{equation}
V_N(x,y) = V(x) + U(y) + v(x,y),
\label{V4}
\end{equation}
where the function $v(x,y)$ describes
the coupling between the two degrees of freedom.
Then the effective potential $V_{{\rm eff}}(x)$
can be presented in the following form,
\begin{equation}
V_{{\rm eff}}(x) = V(x) - T S(x,T),
\label{Veff2}
\end{equation}
where the ``entropy potential'' $S(x,T)$ is 
defined by the expression
\begin{equation}
S(x,T) = \ln \int_{-\infty}^{+\infty} dy \;
  \exp \{ -[U(y) + v(x,y)] /T \}.
\label{Veff3}
\end{equation}
Notice that $S(x)$ does not depend on $V(x)$.

\section{Applications}
\label{appl}

\subsection{Pure entropy barriers}
\label{s3}

\begin{figure}
\epsfxsize=\hsize
\epsfbox{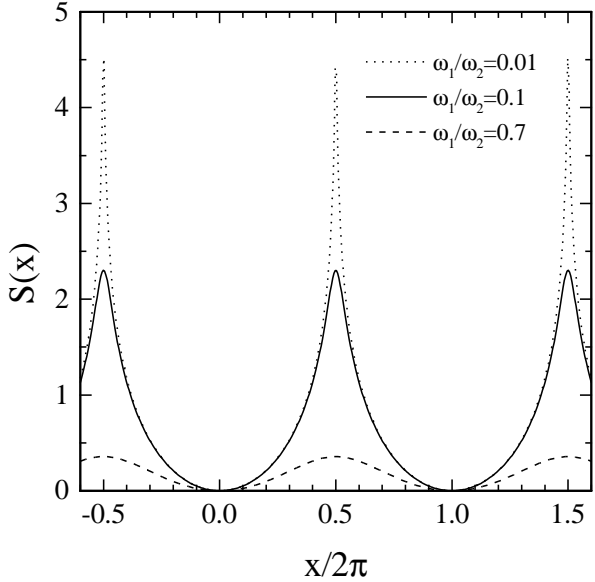}
\caption{
The entropy potential $S(x)$
for pure entropic barriers
with
$\omega_1/\omega_2 =0.01$
(dotted curve),
$\omega_1/\omega_2 =0.1$
(solid curve), and
$\omega_1/\omega_2 =0.7$
(dashed curve), respectively.}
\label{fig01}
\end{figure}

Let $V(x) = 0$ in Eq.\ (\ref{V4}), 
\begin{equation}
U(y) = \frac{1}{2} m \, \omega_1^2 y^2, 
\label{ebr1a}
\end{equation}
and
\begin{equation}
v(x,y) = \frac{1}{4} m \,
(\omega_2^2-\omega_1^2) \, (1-\cos x) \, y^2,
\label{ebr1b}
\end{equation}
so that the atomic motion is inactivated
in the $x$ direction, but
the frequency of transverse oscillation
depends on $x$,
$\omega=\omega_1$ at $x=0$
and $\omega=\omega_2$ at $x=\pi$.
Then the integral in Eq.\ (\ref{Veff3})
can be easily evaluated analytically,
and the entropy potential is given by the expression
\begin{equation}
S(x) = -\frac{1}{2} \ln \left\{ 1
+ \frac{1}{2}
\left[\left(\frac{\omega_2}{\omega_1}\right)^2 -1 \right] (1-\cos x)
\right\}.
\label{ebr3}
\end{equation}
Notice that the entropy potential (\ref{ebr3})
does not depend on temperature, because both potentials
(\ref{ebr1a}) and (\ref{ebr1b}) depend on $y$
in the same way ($\propto y^2$).
The function $S(x)$ is shown in
Fig.\ \ref{fig01}.
It is periodic with the period
$a=2\pi$ and the height
$\varepsilon_S=|\ln (\omega_2/\omega_1)|$.
The diffusion coefficient is given by
$D = D_f F$, where
the entropy factor $F$ depends on 
the ratio of frequencies
$z = \omega_2 / \omega_1$ only,
\begin{equation}
F(z) = [ I_+(z) I_-(z) ]^{-1},
\label{ebr4}
\end{equation}
where
\begin{equation}
I_{\pm}(z) = \pi^{-1} \int_0^{\pi} dx
\left[ 1+\frac{1}{2} (z^2-1) (1-\cos x) \right]^{\pm 1/2}.
\label{ebr5}
\end{equation}
\begin{figure}
\epsfxsize=\hsize
\epsfbox{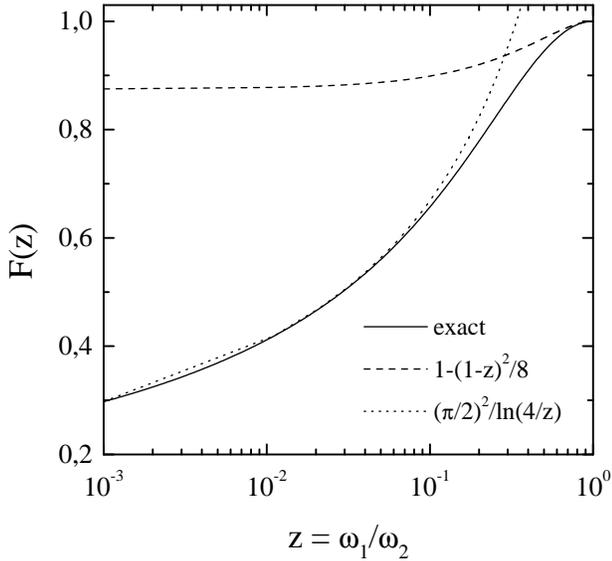}
\caption{
The entropy factor $F(\omega_1/\omega_2)$
for pure entropy barriers
in the overdamped limit.}
\label{fig02}
\end{figure}
Equations (\ref{ebr4},\ref{ebr5}) yield
$F(z)=(\pi/2)^2 \-
{\mathbf K}^{-1} (\sqrt{1-z^2}) \\
{\mathbf E}^{-1} (\sqrt{1-z^2})$,
where ${\mathbf K}$
is the complete elliptic integral of first kind.
Near $z \approx 1$ the function $F(z)$
has the expansion
$F(z) \approx 1- \frac{1}{8} (1-z)^2$,
while at $z \rightarrow 0$ it behaves as
$F(z) \approx (\pi/2)^2 \ln^{-1}(4/z)$.
The function $F(z)$ is presented in Fig.\ \ref{fig02}.
One can see that in the overdamped limit,
the effect of entropy barriers is not too strong,
in particular, even for
$\omega_1/\omega_2 =0.1$
the diffusion coefficient reduces
comparing with the free-diffusion value
by a factor of $F(0.1) \approx 0.66$ only.
Indeed, although the height $\varepsilon_S$
tends to infinity at $z \rightarrow 0$,
the width of barriers becomes very narrow
and thus cannot strongly modify
the diffusion coefficient.

\begin{figure}
\epsfxsize=\hsize
\epsfbox{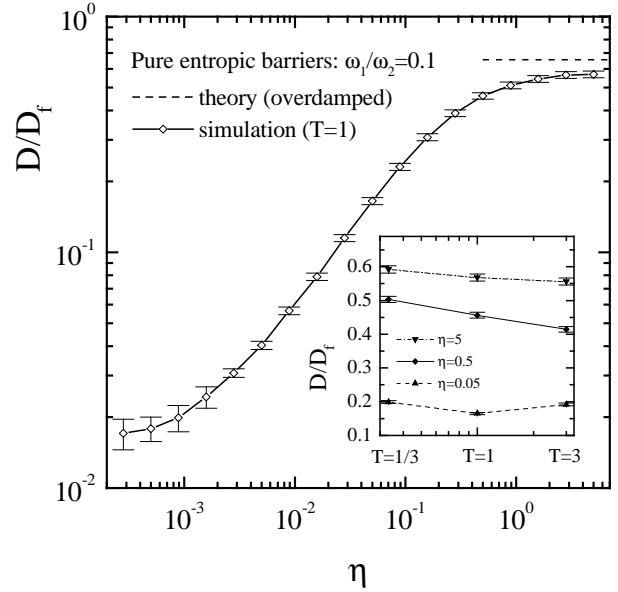}
\caption{
The diffusion coefficient $D/D_f$
(where $D_f=k_B T/m \eta$)
as function of the friction coeffitient $\eta$
for pure entropic barriers
with $\omega_1=0.1$ and $\omega_2=1$ at $T=1$.
Inset: dependence on temperature
($T=1/3$, 1 and 3)
for
$\eta=0.05$ (up triangles), 
0.5 (diamonds) and 
5 (down triangles).}
\label{fig03}
\end{figure}

\begin{figure}
\epsfxsize=\hsize
\epsfbox{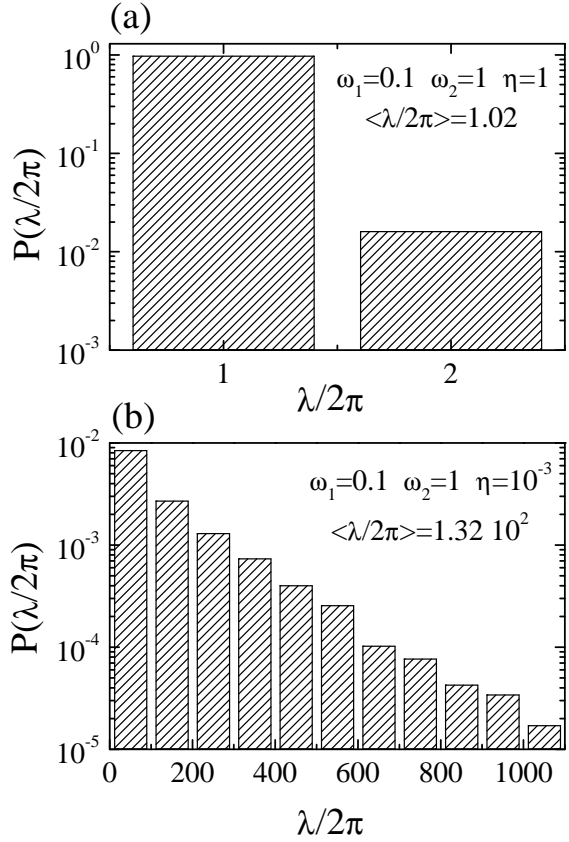}
\caption{
Distribution of jump's length
for pure entropic barriers
with $\omega_1=0.1$ and $\omega_2=1$ at $T=1$
for (a) overdamped case ($\eta=1$) 
and (b) underdamped limit ($\eta=10^{-3}$).}
\label{fig03add}
\end{figure}

In the underdamped case, on the contrary,
the role of entropy barriers is essential.
The dependence of the diffusion coefficient
on the damping constant $\eta$ was obtained
numerically and shown in Fig.\ \ref{fig03}.
One can see that the function $D(\eta)$ exhibits
a typical behavior of activated diffusion
($D \propto \eta^{-1}$
at small and large frictions with a crossover
between the limits) 
as might be expected from the shape of the entropy
potential $S(x)$ of Fig.\ \ref{fig01}.
In the overdamped limit the average jump length is 
equal to the period of the potential $S(x)$,
$\lambda \approx 2\pi$, while
in the underdamped limit
long jumps with $\lambda/2\pi \gg 1$
play the dominant role as shown in 
Fig.\ \ref{fig03add}
(in these simulations we assumed that
the atom is trapped in a given well
if it has sojourned in this well
for a time lapse not shorter than
$(2\eta)^{-1}$ \cite{Risken,Borro}).
The effect of entropy barriers is even stronger than might
be expected from the analogy with the energy
barriers of the same height. For example,
for the frequencies $\omega_1/\omega_2=0.1$
used in the simulation,
the height of the barrier is
$\varepsilon_S=S(\pi) \approx 2.3$,
that would give the ratio
$D(\eta \rightarrow \infty) / D(\eta \rightarrow 0)
\approx 2 \varepsilon_S / k_B T
\approx 4.6$ for the $T=1$ case,
while the simulation leads to the ratio
$D(\eta \rightarrow \infty) / D(\eta \rightarrow 0) > 33$.
From Fig.\ \ref{fig03add}b one can see that
$\langle \lambda/2\pi \rangle \approx 10^2$
for the case of $\eta=10^{-3}$, while
for the one-dimensional diffusion
it should be
$\langle \lambda/2\pi \rangle \sim \eta^{-1} = 10^3$.
Thus, multi-dimensional effects result in a strong
reduction of jump's length in the underdamped
limit which leads to a decrease of the diffusion
coefficient comparing with the 1D motion.
Note also that the dependence on temperature
(shown in inset of Fig.\ \ref{fig03})
is almost negligible as has to be expected
for the entropy potential.
%

\subsection{Atom in a corrugated channel}
\label{s4}

\begin{figure}
\epsfxsize=\hsize
\epsfbox{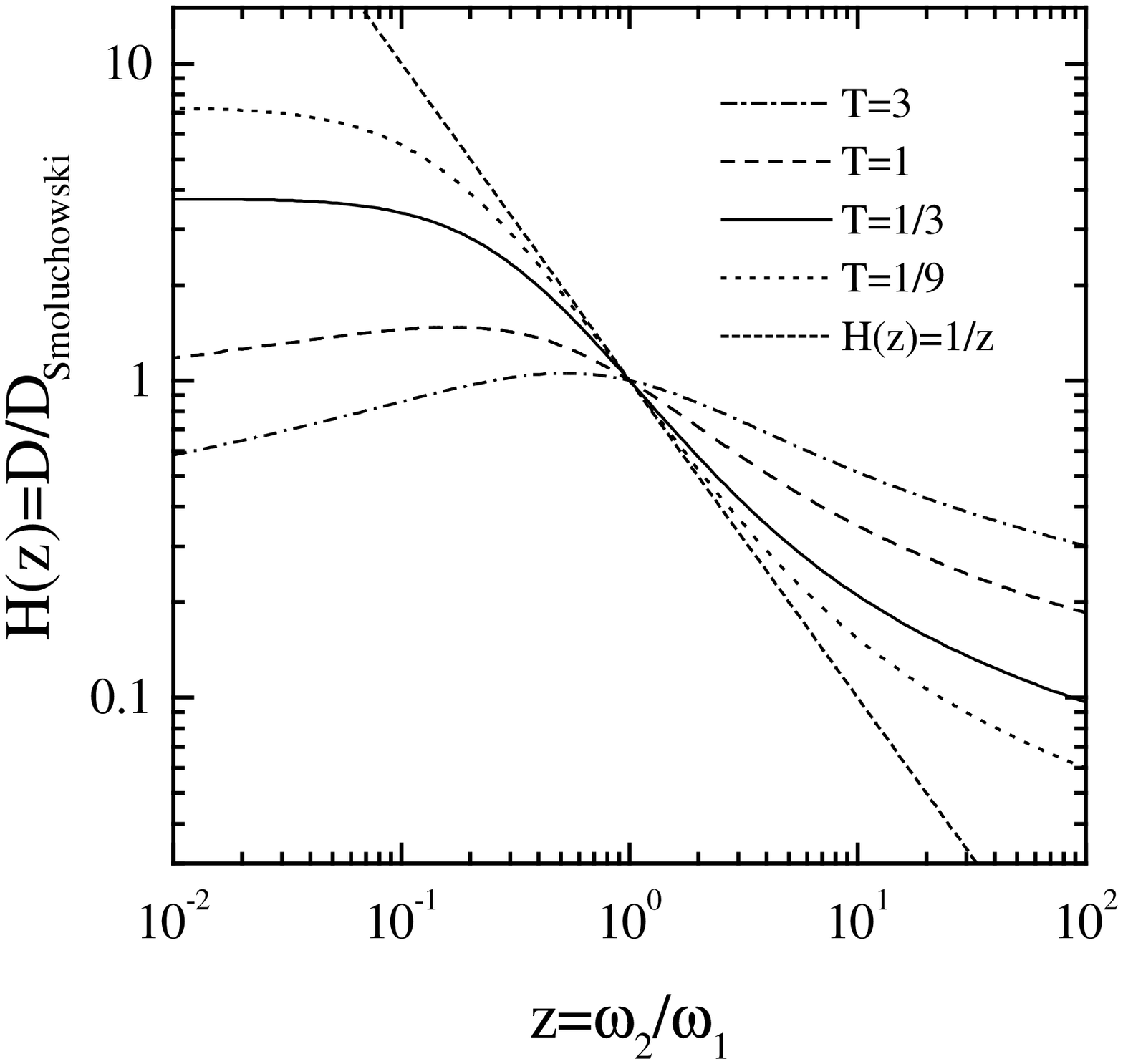}
\caption{
Activated diffusion with the barrier $\varepsilon=2$:
$D$
[normalized on the Smoluchowski value (\ref{DHF})]
versus the ratio of transverse frequencies
$z=\omega_2 / \omega_1$
in the overdamped limit
for the temperatures
$T=3$ (dot-dashed curve),
$T=1$ (dashed curve),
$T=1/3$ (solid curve), and
$T=1/9$ (dotted curve).
The short-dashed line shows the TST approximation.}
\label{fig04}
\end{figure}

\begin{figure}
\epsfxsize=\hsize
\epsfbox{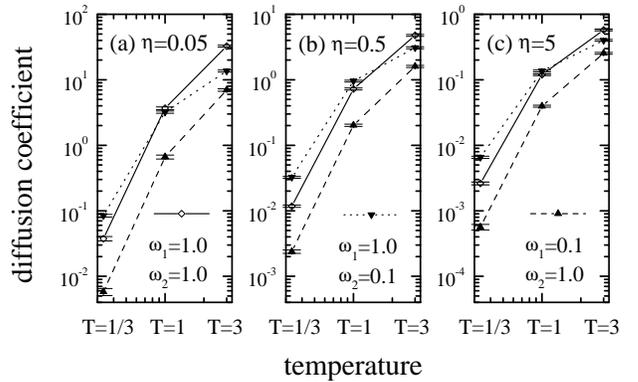}
\caption{
$D$ versus $T$
for the activated motion with the barrier
$\varepsilon=2$
for three values of the transverse frequencies
(open diamonds and solid curves for
$\omega_1=\omega_2=1.0$,
down triangles and dotted curves for ``wide barriers'' with
$\omega_1=1.0$ and $\omega_2=0.1$, and
up triangles and dashed curves for ``narrow barriers'' with
$\omega_1=0.1$ and $\omega_2=1.0$)
for three values of the external damping:
(a) $\eta=0.05$, 
(b) $\eta=0.5$, and
(c) $\eta=5$.}
\label{fig05}
\end{figure}

\begin{figure}
\epsfxsize=\hsize
\epsfbox{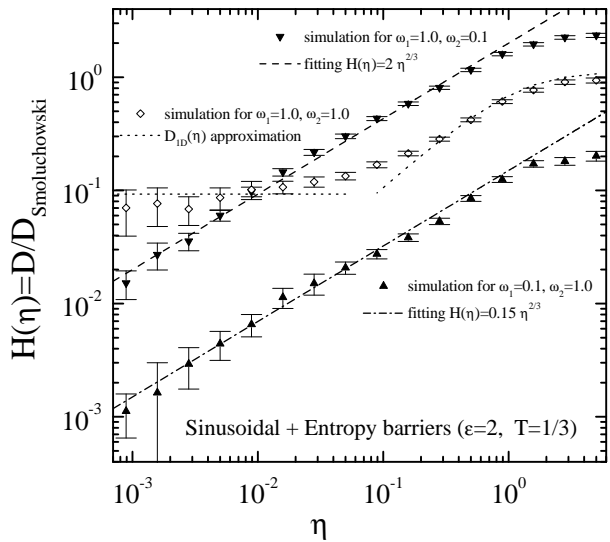}
\caption{
The diffusion coefficient $D$
[normalized on the Smoluchowski value (\ref{DHF})]
as function of the friction coefficient $\eta$
for the activated motion with the barrier
$\varepsilon=2$ at $T=1/3$
for three values of the transverse frequencies:
(a) $\omega_1=\omega_2=1.0$
[open diamonds, the dotted curves show
the 1D approximate values
$D \approx (\omega_0/2\pi) A a^2 B(\eta)$
and
$D \approx \pi D_f A/2$
at high and low frictions respectively],
(b) $\omega_1=1.0$ and $\omega_2=0.1$
(down triangles, ``wide barriers''), and
(c) $\omega_1=0.1$ and $\omega_2=1.0$
(up triangles, ``narrow barriers'').
The dashed curves show the fit
$D(\eta) \propto \eta^{-1/3}$.}
\label{fig06}
\end{figure}

Let now the dependence of the external potential
$V_N(x,y)$
on $y$ is still given by Eqs.\ (\ref{ebr1a}) and (\ref{ebr1b}),
but the motion in the $x$ direction is activated,
\begin{equation}
V(x) = \frac{1}{2} \, \varepsilon \, (1-\cos x),
\label{ads1}
\end{equation}
where $\varepsilon$ is the height of the external potential.
At the minima of the potential (\ref{ads1})
the transverse vibrations are characterized
by the frequency $\omega_1$,
while at the saddle points,
by the frequency $\omega_2$.
In the one-dimensional case,
as well as for the 2D case with $\omega_1 = \omega_2$, 
in the overdamped limit we have,
according to Eq.\ (\ref{DHF}),
$D_{{\rm Smoluchowski}} = D_f \, 
{\rm I}_0^{-2}(\varepsilon / 2 T)$.
Because the entropy potential $S(x)$ does not depend
on the function $V(x)$, it is still given by
Eq.\ (\ref{ebr3}), and the integral (\ref{D0})
can be easily evaluated.
The results for the overdamped limit are shown in Fig.\ \ref{fig04},
which can be compared with
the simulation results for different frictions presented in
Figures \ref{fig05} and \ref{fig06}.
From the $D(T)$ dependence 
of Figures \ref{fig04} and \ref{fig05} one can see that
at high temperatures,
when the motion is inactivated,
the $\omega_1=\omega_2$ case leads to
the maximum of the diffusion coefficient
similarly to the case with pure entropic barriers.
With temperature decreasing,
the energy barriers and the entropy barriers
play ``in phase'' for the ``narrow-barriers'' case of
$\omega_1 < \omega_2$,
and ``in antiphase'' for the ``wide-barriers'' case of
$\omega_1 > \omega_2$.
At low temperatures
$D>D_{\rm 1D}$ for the case of
$\omega_1 > \omega_2$
at high and moderate frictions
in agreement with predictions of the TST approach.
The effect, however, is smaller than
the TST predicts: in simulation we found that
the diffusion coefficient changes
only in three times when the ratio of frequencies
is equal to ten.
At very low frictions
(e.g., $\eta < 10^{-2}$ in Fig.\ \ref{fig06}),
the entropy barriers become more important
than the energy barriers, and
the diffusion coefficient again becomes smaller
than the 1D one for all cases of
$\omega_1 \neq \omega_2$
as it was for the case of pure entropic barriers.
For moderate and low frictions
the simulation results of Fig.\ \ref{fig06}
can be fitted by a dependence
$D(\eta) \propto \eta^{-1/3}$.
Because the escape rate $R$
is still proportional to $\eta$
in the multi-dimensional case
\cite{Borkovec},
we may conclude that in the present case,
the average jump length scales as
$\langle \lambda \rangle \propto \eta^{-2/3}$,
which is in agreement with the results
of pure entropic barriers presented in the
previous subsection,
and also may be compared with the 1D law
$\langle \lambda \rangle \propto \eta^{-1}$
and the 2D simulation result
\cite{Chen}
$\langle \lambda \rangle \propto \eta^{-3/4}$.
Thus, in the underdamped limit multi-dimensional effects
lead to decreasing of diffusivity
(comparing with the 1D case)
due to reduction of jump length which scales as
$\langle \lambda \rangle \propto \eta^{-2/3}$
instead of the 1D scaling law
$\langle \lambda \rangle \propto \eta^{-1}$.

\subsection{Diffusion of the dimer}
\label{s5}

Now we can study diffusion of a dimer in
the 1D sinusoidal potential.
Let $x_1$ and $x_2$ are the coordinates
of two atoms coupled by the elastic spring with the constant $g$,
and $a_0$ is the equilibrium distance
($0 \le a_0 \le \pi$).
Then the Hamiltonian of the system takes the form
\begin{eqnarray}
H =
\frac{1}{2} m_a \dot{x}_1^2 +
\frac{1}{2} m_a \dot{x}_2^2 +
\frac{1}{2} \varepsilon_s (1-\cos 2\pi x_1 /a_s) +
\nonumber \\
\frac{1}{2} \varepsilon_s (1-\cos 2\pi x_2 /a_s) +
\frac{1}{2} \, g \, (x_2-x_1-a_0)^2.
\label{dim1}
\end{eqnarray}
In what follows we put $\varepsilon_s=2$, $m_a=1$, $a_s=2\pi$,
and in the present paper we consider the case of
$a_0=0$ only.
Introducing the coordinates
$x = x_1+x_2$ and $y = x_2-x_1$,
the Hamiltonian (\ref{dim1}) can be rewritten as
\begin{equation}
H =
\frac{1}{2} m \left( \dot{x}^2 + \dot{y}^2 \right) +
V_N (x,y),
\end{equation}
\begin{equation}
V_N (x,y) =
\frac{1}{2} \varepsilon \left( 
1-\cos \frac{x}{2} \; \cos \frac{y}{2} \right)
+\frac{1}{2} \, g \, y^2,
\label{dim2}
\end{equation}
which describes the motion of one particle of mass
$m=m_a/2=1/2$
in the $x$-periodic potential of height
$\varepsilon=2\varepsilon_a=4$
and period $a=2 a_s=4 \pi$.

\begin{figure}
\epsfxsize=\hsize
\epsfbox{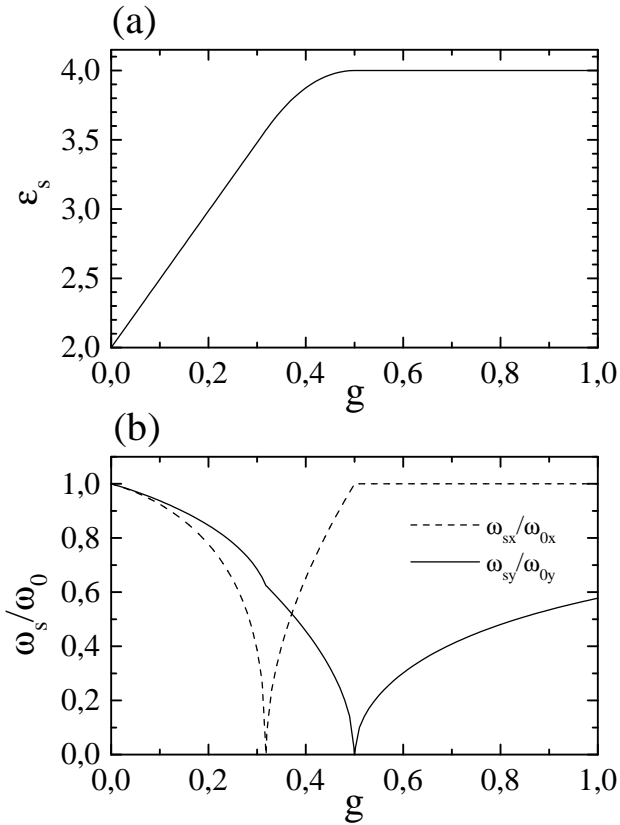}
\caption{
(a) The activation energy $\varepsilon_s$ and
(b) the ratio of frequencies at the saddle
and minimum points
as functions of the elastic constant $g$
for dimer's diffusion.}
\label{fig07}
\end{figure}

\begin{figure}
\epsfxsize=\hsize
\epsfbox{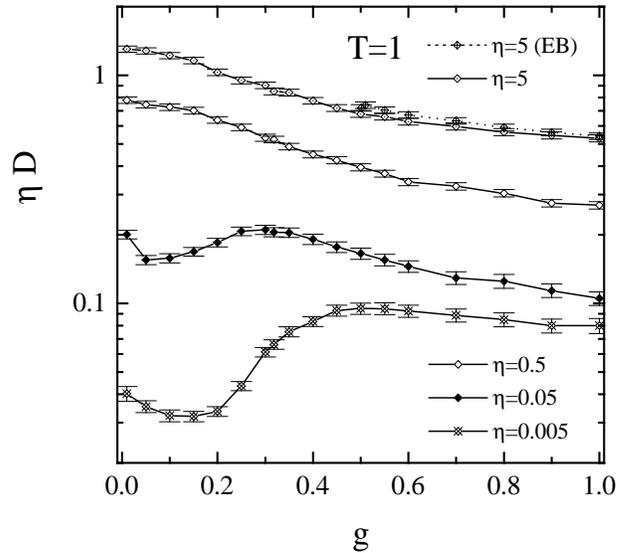}
\caption{
The dependence of the diffusion coefficient $D$
(times $\eta$)
on the elastic constant $g$ at $T=1$
for different values of the damping constant:
$\eta=5$ (dotted diamonds),
$\eta=0.5$ (open diamonds),
$\eta=0.05$ (solid diamonds), and
$\eta=0.005$ (crossed diamonds).
The dotted curve and plussed diamonds
show the simulation results for the 
``atom in channel'' model with $\eta=5$ and other
parameters adjusted to the dimer case.}
\label{fig08}
\end{figure}

\begin{figure}
\epsfxsize=\hsize
\epsfbox{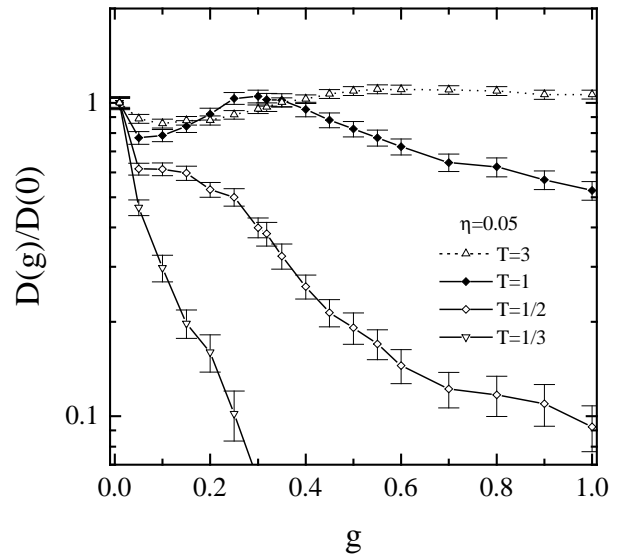}
\caption{
The ration $D(g)/D(0)$ as function
of the elastic constant $g$
for the dimer diffusion at $\eta=0.05$ and
different temperatures $T=3, 1, 1/2$, and $1/3$.}
\label{fig09}
\end{figure}

The adiabatic trajectory for this system
was studied in \cite{SSmine}.
Its shape depends on a value
of the elastic constant $g$.
The points $(x,y)=(4\pi n,0)$,
where $n$ is an integer, always correspond to the
absolute minimum of the potential energy.
Near the minimum, the potential energy has
the expansion
$V_N (x,y) \approx \frac{1}{2} m \left(
\omega_{0x}^2 x^2 + \omega_{0y}^2 y^2 \right)$
with
$\omega_{0x} = 1$ and
$\omega_{0y} = (2g+1)^{1/2}$.
For a strong spring, $g \geq 1/2$,
there is only one saddle point at 
$(x_s,y_s)=(2\pi,0)$
between two adjacent minima $(0,0)$ and $(4\pi,0)$.
Near the saddle, the potential energy has
the expansion
\begin{equation}
V_N (x,y) \approx \varepsilon_s + \frac{1}{2} m \left[
- \omega_{sx}^2 (x-x_s)^2 
+ \omega_{sy}^2 (y-y_s)^2 \right]
\label{expan}
\end{equation}
with
$\omega_{sx} = 1$ and
$\omega_{sy} = (2g-1)^{1/2}$,
so that the activation energy for dimer motion
is equal to $\varepsilon_s = \varepsilon = 4$
(see Fig.\ \ref{fig07}).
Therefore, dimer diffusion can be approximately described
as motion of one atom in the corrugated periodic
potential with the transverse frequencies
$\omega_{1,2}=(2g \pm 1)^{1/2}$,
i.e.\ it corresponds to the case of ``wide'' 
barriers studied in the previous subsection.
Thus, although the shape of adiabatic trajectory
does not depend on the elastic constant for the
case of strong coupling,
the diffusion coefficient does depend on $g$,
it increases when $g \rightarrow 1/2$ 
due to decreasing of
the transverse curvature at the saddle point.
The simulation results of Fig.\ \ref{fig08}
show that the harmonic approximation
describes the $D(g)$ dependence with a good accuracy.
From Fig.\ \ref{fig09}, where the ratio
$D(g)/D(0)$ is presented for different temperatures,
one can see also that close to the critical
point $g=1/2$, when anharmonicity of transverse
vibrations at the saddle point is large,
the entropy factor strongly depends on $T$,
especially at low temperatures.

For intermediate values of the elastic constant,
$1/ \pi \leq g < 1/2$,
the adiabatic trajectory still has only one
saddle point $(2\pi,y_s)$
between the adjacent minima,
where $y_s$ is now a solution of the transcendental equation
$\sin (y_s /2) = g y_s$.
Near the saddle, the potential energy has
the expansion (\ref{expan})
with the frequencies
$\omega_{sx} = \left[ 1-(g y_s)^2 \right]^{1/4}$ and
$\omega_{sy} = (2g-\omega_{sx}^2)^{1/2}$.
The saddle is characterized by the energy
$\varepsilon_s (g) = \frac{1}{2} \varepsilon
\left[ 1+ \cos (y_s /2) \right]
+ \frac{1}{2} g y_s^2$,
so that
$2 + \pi /2 < \varepsilon_s < 4$.


Finally, for a weak coupling between dimer's atoms,
$g < 1/ \pi$,
there are two saddle points
between the adjacent minima $(0,0)$ and $(4\pi,0)$,
with a local minimum of the potential energy
between these saddle points.
The coordinates of the saddle points are
$(2\pi-x',\pi)$ and $(2\pi+x',\pi)$,
where $x'=2 \cos^{-1} (g \pi)$.
These saddle points are
characterized by the energy
$\varepsilon_s (g) = \frac{1}{2} ( \varepsilon + g \pi^2)$,
so that
$2 < \varepsilon_s < 2 + \pi /2$.
Near the saddle, the potential energy has
the expansion (\ref{expan})
with coefficients
$\omega_{sx} = (g-G)^{1/2}$ and
$\omega_{sy} = (g+G)^{1/2}$,
where $G=\left[1- (\pi^2-1)g^2 \right]^{1/2}$.

The whole dependence $\varepsilon_s(g)$ 
is shown in Fig.\ \ref{fig07}a.
The activation energy monotonically increases
from the single-atom value $\varepsilon_s=2$
at $g=0$ to the rigid-dimer value $\varepsilon_s=4$
at $g=1/2$ and then remains constant.
Thus, one could expect that the diffusion coefficient
should monotonically decrease with $g$ increasing.
However, the simulation results of Fig.\ \ref{fig09}
show that often this is not true.
The peculiarity in the transverse frequencies
at the point $g=1/2$,
where the saddle transverse frequency reaches zero,
leads to a maximum of the function $D(g)$
close to this point,
if the damping is small,
$\eta \alt 0.5$,
and the temperature is not too low,
$T \agt 1$
(recall $\varepsilon =4$).
Thus, multi-dimensional effects may strongly
affect dimer's diffusivity.

\section{Conclusion}
\label{s6}

In the present paper we studied in details
the diffusion of a particle in two-dimensional
space which is periodic along $x$ and unbounded
in the transverse direction.
We calculated the entropy factor
which emerges due to transverse degree of freedom,
both in the overdamped limit (analytically)
and in the underdamped case (numerically),
and compared it with the prediction
of the transition-state theory.
We showed that
in the underdamped limit, the multi-dimensional effects
lead to reduction (comparing with the one-dimensional motion)
of jump lengths between subsequent trapping
of the atom in bottoms
of the external periodic potential.
The simulation predicts that jump lengths scale as
$\langle \lambda \rangle / \langle \lambda_{\rm 1D} \rangle 
\propto \eta^{1/3}$.
This leads to a decrease of diffusivity
which now scales as $D \propto \eta^{-1/3}$
instead of the 1D dependence $D_{\rm 1D} \propto \eta^{-1}$.

In the overdamped limit, the entropy factor
(and, therefore, the prefactor in the Arrhenius
formula for activated diffusion)
does not depend on temperature
as long as the transverse motion
near the adiabatic trajectory may be described
by the harmonic approximation.
Simulation shows that this remains true,
at least approximately, for low damping as well.
Thus, in most cases the experimentally observed dependence
of the prefactor on temperature has to be
attributed to collective effects due to
interaction between diffusing particles
or/and between the atom and (deformable) substrate.
However, in the case of dimer diffusion at some value
of the interaction between the atoms,
when the saddle transverse frequency is equal zero,
the anharmonicity of the transverse potential
begins to play the important role and
the entropy factor strongly depends on $T$.


\acknowledgments
Helpful discussions with T.\ P.\ Valkering are
gratefully acknowledged.
This work was partially supported by
the INTAS Grant 97-31061.


\end{document}